# Universal Magnetic-Field-Driven Metal-Insulator-Metal Transformations in Graphite and Bismuth

Y. Kopelevich, J. C. Medina Pantoja, R. R. da Silva, and S. Moehlecke

Instituto de Física "Gleb Wataghin", Universidade Estadual de Campinas, Unicamp 13083-970, Campinas, São Paulo, Brasil

ABSTRACT

Applied magnetic field induces metal-insulator and re-entrant insulator-metal transitions in both graphite and rhombohedral bismuth. The corresponding transition boundaries plotted on the magnetic field - temperature (B - T) plane nearly coincide for these semimetals and can be best described by power laws $T \sim (B - B_c)^\kappa$, where $B_c$ is a critical field at $T = 0$ and $\kappa = 0.45 \pm 0.05$. We show that insulator-metal-insulator (I-M-I) transformations take place in the Landau level quantization regime and illustrate how the IMT in quasi-3D graphite transforms into a cascade of I-M-I transitions, related to the quantum Hall effect in quasi-2D graphite samples. We discuss the possible coupling of superconducting and excitonic correlations with the observed phenomena, as well as the signatures of quantum phase transitions associated with the M-I and I-M transformations.

The behavior of matter in strong magnetic field is of fundamental and broad interest ranging from quantum electro- and chromo-dynamics [1-3] to the condensed matter physics [4]. The strong enough magnetic field (B ≥ $B_{QL}$) applied to a bulk conductor pulls carriers into the lowest Landau level (LLL) which reduces their effective dimensionality from 3D to 1D and enhances fermion pairing instabilities towards an excitonic insulator [5], superconductor [6], spin- and charge-density wave (SDW and CDW) states [7, 8], or Luttinger liquid state [9]. A magnetic-field-induced unconventional Fermi liquid which differs from both the Luttinger- and Landau-type has also been proposed [10].

Graphite and bismuth (Bi) are considered to be promising materials for experimental explorations of the above phenomena. These semimetals possess low densities of electrons (e) and holes (h); $N_e \sim N_h \sim 10^{17}$ cm$^{-3}$ in Bi [11], and $N_e \sim N_h \sim 10^{17}... 10^{18}$ cm$^{-3}$ in graphite [12] with very small effective masses; m$^*$ ~ $10^{-3} – 10^{-2} m_0$ [11, 12], $m_0$ being the free electron mass. For light carriers, the quantum limit, for instance in graphite, can be reached at B ≥ $\mu_0$H ~ 0.1 T, and B ≡ $B_{QL}$ ≈ 7 T pulls all carriers into LLL [12]. The magnetic-field-driven CDW-like state has been observed in magnetoresistance measurements [13, 14] for single crystalline graphite at B ≥ 20 T, i. e. in the ultra-quantum limit. Indications for an excitonic phase in the quantum limit have been found in acoustic measurements of bismuth [15].

Recently, both magnetic-field-driven metal-insulator and reentrant insulator-metal transformations (MIT and IMT) have been measured in these semimetals at much lower fields (~ 0.1…1 T) by various groups [16-18]. The obtained results have been analyzed in terms of excitonic and superconducting instabilities [16] or using classical multi-band models [17, 18]. Also, the quantum Hall effect (QHE) has been observed in strongly anisotropic quasi-2D graphite and graphene (single layer of graphite) [16, 19-21], accompanied by field-induced cascade of insulator-metal-insulator (I-M-I) transitions [16, 21]. Those results together with the identification of Dirac fermions in graphite (graphene) [20-22] provide strong arguments against a classical treatment of the magnetotransport in graphite.

In particular, it is expected [23, 24] that electron-electron interaction and/or applied magnetic field induce the excitonic insulating state in graphite. Thus, the magnetic-field-driven metal-excitonic insulator transition in graphite is seen as the condensed-matter realization of the magnetic catalysis (MC) phenomenon known in relativistic theories of (2+1)-dimensional Dirac fermions [1].

The very small effective mass of carriers in bismuth (~ $0.001m_0$ [11]) implies that these may be considered as Dirac-like fermions in 3+1 dimensions. Because of the effective dimensional reduction in the quantized field, MC phenomenon is expected to occur in 3+1 dimensions as well [2]. Hence, the underlying physics in bismuth and graphite may not be so different. The reduced dimensionality of bismuth (3D → 2D) due to surface effects [25, 26] suggests even more close analogy between these materials.

If the Landau level quantization and the related quantum phenomena dominate the metal-insulator and insulator-metal transformations, one expects a quite similar or even universal (magneto)transport properties of bismuth and graphite in spite of their quite different electronic band structures. Hence, a comparative study of these materials can be a "smoking gun" proof of one or another approach.

Here we present the results of magnetoresistance and Hall effect measurements performed on rhombohedral Bi and graphite. The results demonstrate the occurrence of magnetic-field-driven MIT and reentrant insulator-metal transitions in both materials. The corresponding transition boundaries plotted on the magnetic field - temperature (B - T) plane nearly coincide and can be best described by dependencies $T \sim (B - B_c)^\kappa$ with $\kappa = 0.45 \pm 0.05$. Such power laws usually appear in the scaling theory of quantum phase transitions (QPT) [27]. In our case, this would imply the existence of two zero-temperature critical fields $B_c^{IMT} > B_c^{MIT}$. On the other hand, it is also found that the two-parameter scaling analysis proposed by Das and Doniach [28] to characterize the Bose metal (the nonsuperconducting state of Cooper pairs) - insulator transition (BMIT) in 2D systems can be well applied to the MIT measured in graphite and Bi, providing another indication for a deep similarity of the physical processes operating in these systems.

Measurements of longitudinal $\rho(B,T)$ and transverse (Hall) $\rho_H(B,T)$ resistivities were performed on several oriented polycrystalline samples of rhombohedral (A7) bismuth, consisting of single crystalline blocks of size $\sim 1 \times 1$ mm$^2$ in the plane perpendicular to the trigonal c-axis. The measured sample resistivity $\rho(T = 300$ K, $B = 0) = 150$ $\mu\Omega$cm and the Hall constant $R_H(T = 2$ K, $B = 0.01$T$) = -1.55\cdot10^{-5}$ m$^3$/C are in a good agreement with the values quoted in the literature [11]. Low-frequency (f = 1 Hz) and dc magnetotransport measurements were performed using standard four-probe as well as van der Pauw methods in the temperature interval 2 K $\leq$ T $\leq$ 300 K by means of PPMS (Quantum Design) and Janis 9T-magnet He-cryostats. Complementary magnetization measurements M(B, T) were carried out with a commercial SQUID magnetometer (Quantum Design). Three Bi samples with dimensions 5.85 x 5.75 x 0.08 mm$^3$ (Bi-S1) and 5.4 x 0.31 x 0.2 mm$^3$ (Bi-S2) and 4 x 3 x 0.5 mm$^3$ (Bi-S3) obtained from the same bar were used in transport and magnetic measurements. Single crystalline (Kish) and polycrystalline highly oriented pyrolitic graphite (HOPG) samples studied in this work have been thoroughly characterized elsewhere; see Ref. [16] and references therein. Briefly, for B = 0 and at room temperature the HOPG samples have an out-of-plane/basal-plane resistivity ratio $\rho_c/\rho_b = 8.6\times10^3$ (HOPG-1) and $\sim 5\times10^4$ (HOPG-3 and HOPG-UC), and the Kish single crystal (Kish) presents the ratio of $\sim 100$; $\rho_b$-values at T = 300 K (B = 0) are $\sim 3$ $\mu\Omega\cdot$cm (HOPG-UC), $\sim 5$ $\mu\Omega\cdot$cm (HOPG-3, Kish), and $\sim 45$ $\mu\Omega\cdot$cm (HOPG-1). From x-ray rocking curve measurements the following values of FWHM (full width at half maximum) were obtained: 0.3° (HOPG-UC), 0.5° (HOPG-3), 1.4° (HOPG-1), and 1.6° (Kish).

Figure 1 presents the longitudinal resistivity ρ(T) obtained on the Bi-S1 sample at various magnetic fields applied perpendicular to the main sample surface (H ∥ c). This figure illustrates the field-induced suppression of the metallic (dρ/dT > 0) state occurring at T < $T_{min}$(B) and its re-appearance at T < $T_{max}$(B) for higher fields.

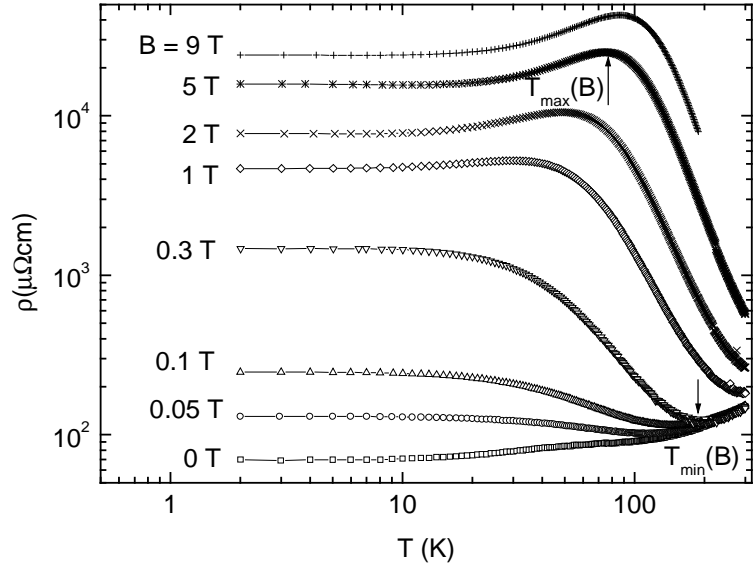

Figure 1. Resistivity measured for Bi-S1 sample for various applied magnetic fields B ∥ ⊥ I. $T_{min}$(B) and $T_{max}$(B) mark MIT and IMT temperatures, respectively.

Figure 2 gives the reduced resistivity ρ(T)/ρ(T = 50 K) curves measured in the low-field (0 ≤ B ≤ 0.05 T) limit. Fig. 1 and Fig. 2 demonstrate also that both $T_{min}$(B) and $T_{max}$(B) are increasing functions of the field. As follows from Fig. 2, the minimum in ρ(T) becomes visible at $T_{min}$ = 91 K for B = 0.03 T, and it rapidly develops with a further field increasing such that for B = 0.05 T the insulating-type (dρ/dT < 0) resistance behavior takes place at T ≤ $T_{min}$ = 116 K.

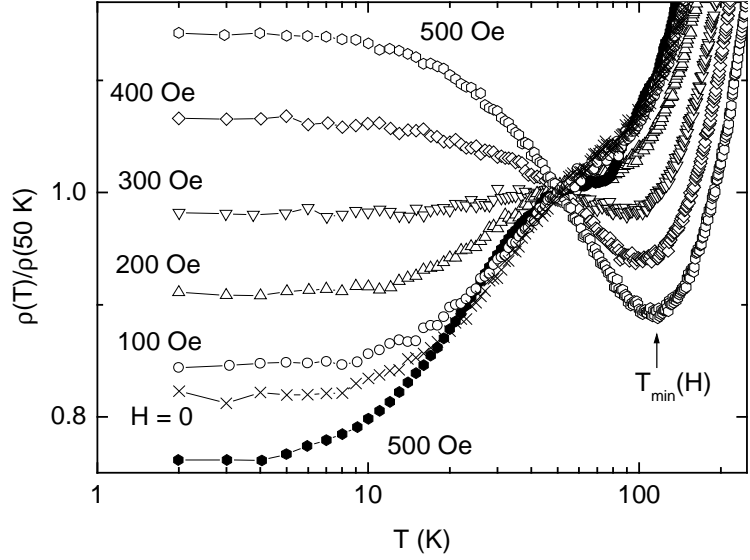

Figure 2. Reduced resistivity ρ(T)/ρ(50 K) obtained for Bi-S1 sample at zero and low applied magnetic fields (open symbols) in configuration B ∥ c ⊥ I. Solid symbols corresponds to the reduced resistivity measured for Bi-S2 sample with B ∥ c ∥ I. $T_{min}(H)$ is the same as in figure 1.

At the same time, no transition to the insulating state was observed in a specially designed sample geometry (Bi-S2) with B = 0.05 T applied parallel to the current (B ∥ I, see Fig. 2). In general, magnetoresistance arises from both bending of the electron trajectory in magnetic field (orbital effect) and spin-dependent scattering mechanism(s). In the Lorentz-force-free configuration, used in the Bi-S2 sample measurements, the classical MR due to bending of electron trajectory is strongly suppressed but spin-scattering channel should not be affected. Then, the non-observation of a MIT in this configuration provides evidence that similar to graphite [16] MIT in bismuth is governed by orbital and not spin-related effects.

The results obtained for the Bi-S1 sample are summarized in Fig. 3 where $T_{min}(B)$ and $T_{max}(B)$ measured for the single crystalline Kish graphite [16, 29] are also plotted.

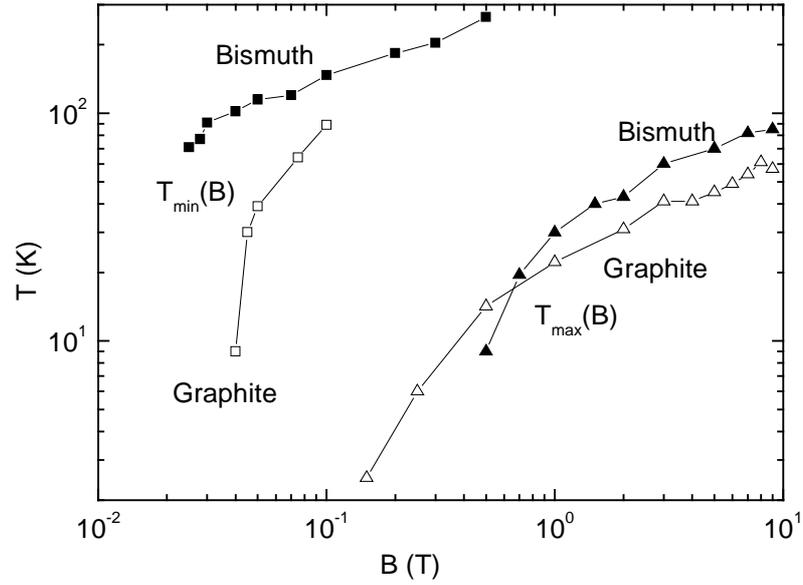

Figure 3. Temperature (T) – applied magnetic field (B) diagram. $T_{min}(B)$ and $T_{max}(B)$ are MIT and IMT transition boundaries, respectively, obtained for both bismuth (Bi-S1) and single crystalline Kish graphite samples.

As can be seen from Fig. 3, $T_{max}(B)$ obtained for Bi and graphite nearly coincide, and $T_{min}(B)$ curves are very close to each other. The striking universality of the magnetotransport in these semimetals can be appreciated from Fig. 4, which illustrates that the curves presented in Fig. 3 can be very well described by $T_{min}(B) \sim (B - B_c^{MIT})^\kappa$ and $T_{max}(B) \sim (B - B_c^{IMT})^\kappa$ dependencies for both materials, where $\kappa = 0.45 \pm 0.05$.

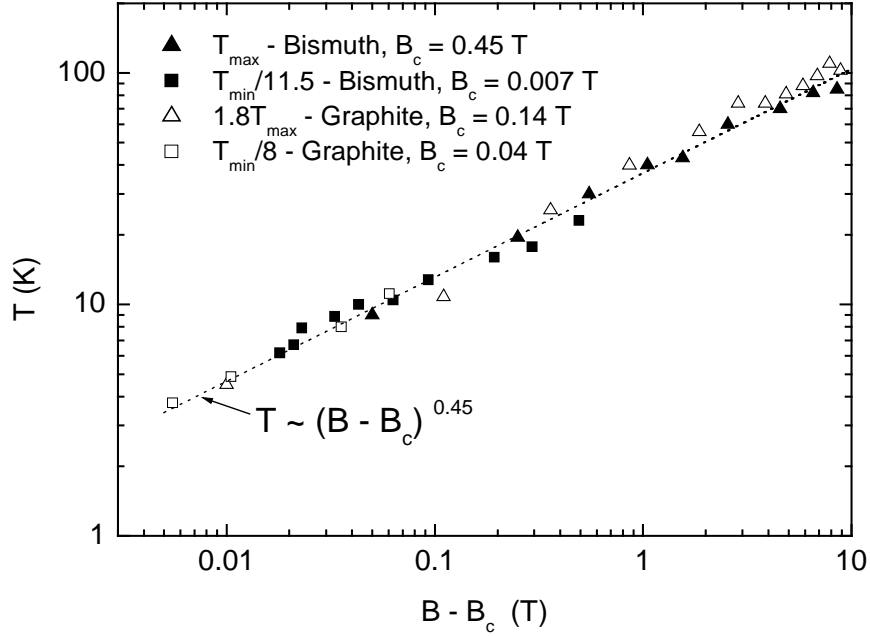

Figure. 4. $T_{min}(B)$ and $T_{max}(B)$ data of Fig. 3 replotted as shown. Dotted line is the power law fit.

Noting, $T_{min}(B)$ and $T_{max}(B)$ resemble very much boundaries on the B-T plane occurring in a vicinity of magnetic-field-induced quantum critical points (QCP) [30-32]. Thus, $B_c^{MIT}$ and $B_c^{IMT}$ may well represent the zero-temperature critical fields for the metal-insulator and insulator-metal transitions, respectively.

On the other hand, it has been demonstrated in Ref. [33, 34] that MIT in graphite can be formally described in terms of the phenomenological two-parameter scaling approach introduced by Das and Doniach within the context of the BMIT theory [28], which assumes the existence of non-superfluid liquid of Cooper pairs (Bose metal) in the zero-temperature limit. Previous to this, a scaling theory of the superconductor-insulator transition (SIT) has been devoloped by Fisher [35].

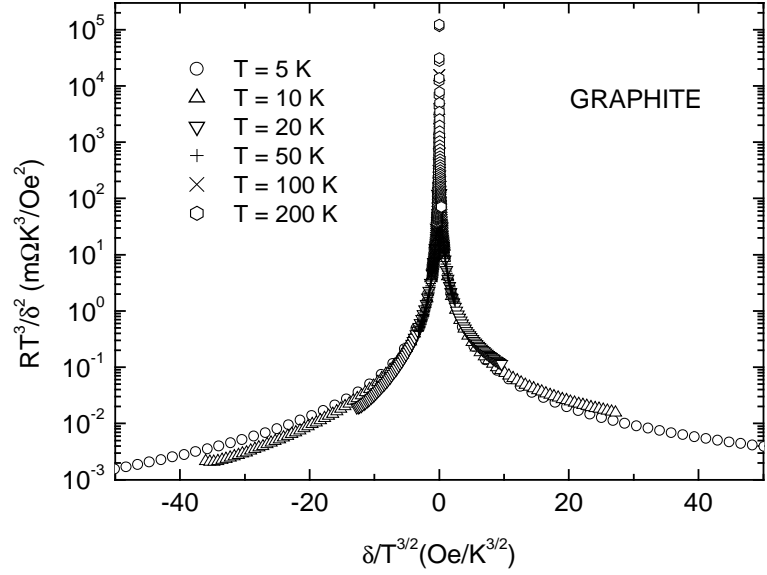

Figure 5. Bose metal – insulator transition scaling analysis of the basal-plane resistance $R = R_b(T, H)$ measured in graphite (HOPG-3); $\delta = H - H_{cr}$, $H_{cr} = 1140$ Oe, $z = 1$, and $\nu = 2/3$.

According to Ref. [35], the resistance in the critical regime of the zero-temperature SIT is given by the equation $R(\delta, T) = R_{cr} f(|\delta|/T^{1/z\nu})$, where $R_{cr}$ is the resistance at the transition, $f(|\delta|/T^{1/z\nu})$ a scaling function such that $f(0) = 1$; $z$ and $\nu$ are critical exponents, and $\delta = H - H_{cr}$ the deviation of the variable parameter (applied magnetic field) from its critical value. However, at low enough temperatures the resistance $R_b(T)$ saturates, and a clear deviation from the scaling takes place in both superconducting films [36] and graphite [29, 33, 34].

At the same time, the two parameter scaling formula $RT^{1+2/z}/\delta^{2\beta} = f(\delta/T^{1/z\nu})$, where $\beta = \nu(z + 2)/2$, proposed in Ref. [28] works very well at low temperatures.

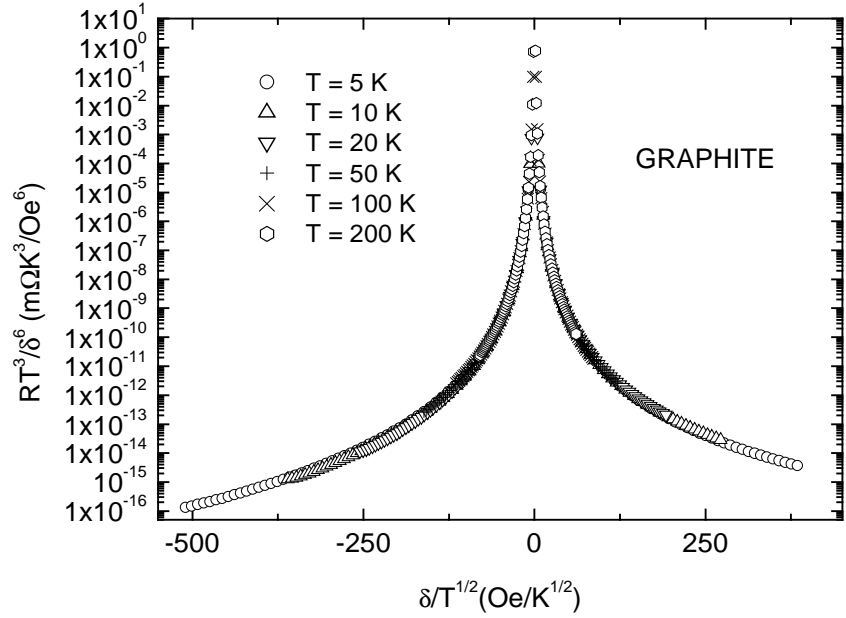

Figure 6. Data of figure 5 replotted with $z = 1$ and $\nu = 2$.

In particular, for HOPG-3 sample, the critical exponents $z = 1$ and $\nu = 2/3$ have been reported [33, 34]. In fact, for magnetic fields not too far from $H_{cr}$, an equally good scaling could be found using $\nu = 4/3$, as reported for Mo-Ge films [36, 28]. For large enough $H - H_{cr}$ the scaling fails, see in Fig. 5 the results for graphite. However, as shown in Fig. 6, the scaling can be recovered taking $z = 1$ and $\nu = 2$ as the critical exponents. Similarly, for bismuth the best scaling is obtained taking $z = 1$ and $\nu = 2$, see Fig. 7.

Before we proceed further with the discussion of possible quantum phase transitions in Bi and graphite, let us see why conventional multiband models [17, 18] cannot be applied. In the first place, the classical approach requires that the Landau level quantization is irrelevant. However, quantum oscillations at low magnetic fields ($H \sim H_{cr}$) are seen in graphite up to 300 K, i. e. for all studied temperatures [37].

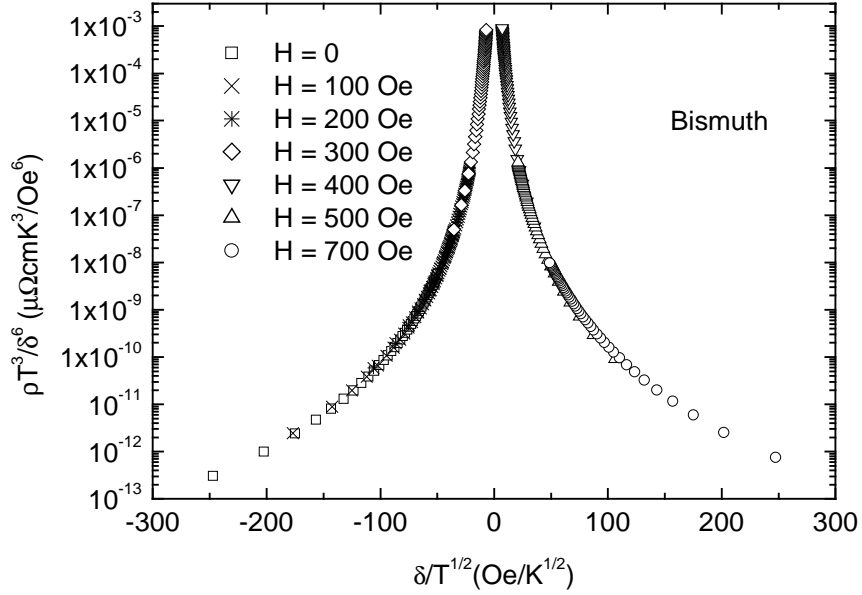

Figure 7. Bose metal – insulator transition scaling analysis of the resistivity $\rho(T, H)$ measured in bismuth (Bi-S1) in geometry ($H \parallel c \perp I$) ; $\delta = H - H_{cr}$, $H_{cr} = 350$ Oe, $z = 1$, and $\nu = 2$.

Figure 8 provides an experimental evidence for Shubnikov-de Haas (SdH) oscillations which take place in bismuth at magnetic fields as low as ~ 30 Oe. For massive carriers, the crossover temperature which separates quantum and classical regimes can be estimated as

$$T_{cr} = heB/2\pi k_B m^*. \qquad (1)$$

Taking $B = 0.003$ T and $T = 2$ K one gets from Eq. (1) $m^* \approx 0.002 m_0$ which is in excellent agreement with previous reports [11].

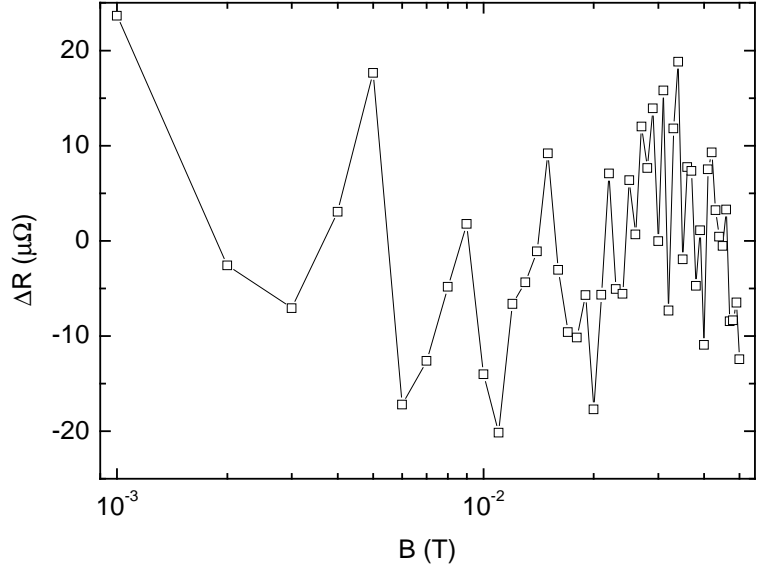

Figure 8. Shubnikov-de Haas oscillations measured for Bi-S2 at T = 2 K, after subtraction of the large polynomial background $R_{bgr}(B)$ signal.

It should be admitted, however, that the oscillating part of the measured resistance is rather small in both Bi and graphite, and one can question the importance of the Landau level quantization in the context of MIT discussion. In other words, one has to verify whether MR can still be described within the frameworks of classical Fermi liquid models or not.

We analyzed carefully the measured MR across the MIT boundary and found that the classical result for orbital MR, i. e. $\Delta R(B)/R \sim B^n$ with n = 2 is violated in the "insulating" regime. Figure 9 which presents MR obtained for both Bi and graphite samples at T = 5 K demonstrates that the parabolic MR occurs only at very low fields (B < $B_{MIT}$), whereas n = 1.25 (graphite) and n = 1.35 (bismuth) are found for B > $B_{MIT}$.

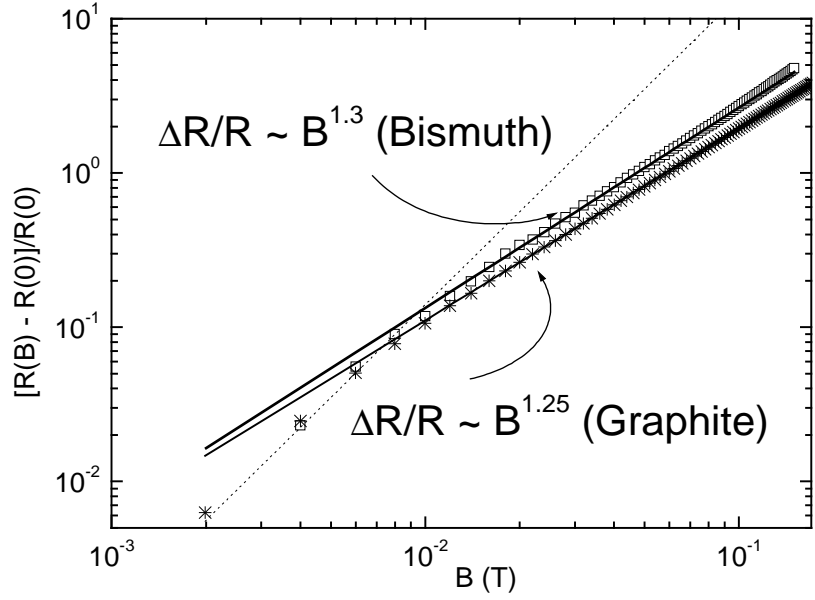

Figure 9. Reduced magnetoresistance $\Delta R(B)/R = [R(B) - R(0)]/R(0)$ measured at T = 5 K for graphite (HOPG-UC) and bismuth (Bi-S1) samples. Bold solid lines are power law fits to the data as indicated in the figure; dotted line corresponds to parabolic magnetoresistance ($\Delta R/R \sim B^2$).

We stress that the obtained exponent values are not consistent with those (n = 1 or n = 2) expected for Fermi liquids [38] but agree with the exponent values found e. g. in the organic conductor $(TMTSF)_2PF_6$ (n = 1.25 - 1.5) [39, 40]. The exponent n < 2 has also been measured in underdoped superconducting cuprates in the pseudogap state [41]; both $(TMTSF)_2PF_6$ and underdoped cuprates are widely considered as non-Fermi liquid systems. Our results obtained for several graphite samples show that deviation from the Fermi liquid behavior (n = 2) takes place precisely at the $T_{min}(B)$ boundary. Figure 10 illustrates this fact where the data obtained for HOPG-UC sample [29] are given.

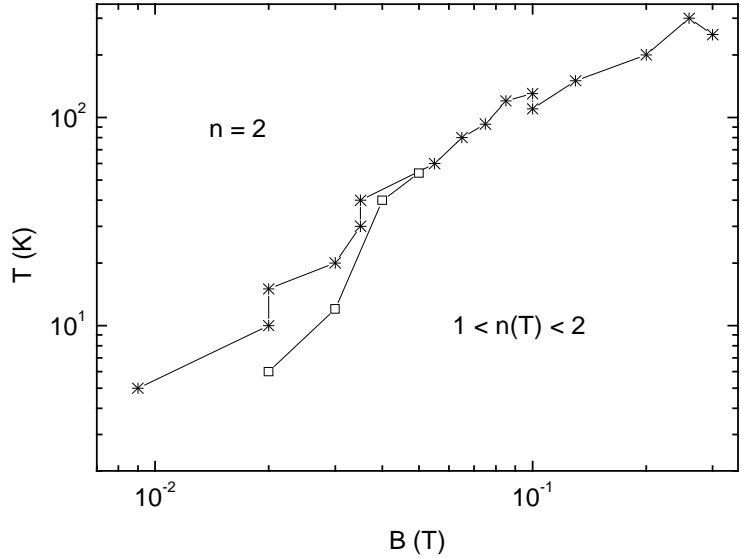

Figure 10. Data points (*) which separate parabolic (n = 2) and non-parabolic, 1 < n(T) < 2, magnetoresistance ($\Delta R/R \sim B^n$) measured for HOPG-UC sample; ( ) - $T_{min}(B)$ data obtained for the same sample [29].

The Hall effect measurements provide an additional evidence for non-classical behavior of graphite and bismuth. Figure 11 depicts the Hall resistivity $\rho_H(B)$ measured for HOPG-2 and HOPG-3 samples at T = 5 K. One can see that for low enough fields $\rho_H$ is a non-linear function of the field, as also observed in Ref. [18]. Besides, measurements performed on our most disordered HOPG sample (HOPG-1) revealed a small but well defined hysteresis in $\rho_H(B)$. As Fig. 12 demonstrates, the hysteresis is clear for - 0.05 T < B < 0.05 T. It is also found that both the hysteresis and the non-linearity in $\rho_H(B)$ vanish with temperature above approximately 150 K. As shown in the inset in Fig. 11, the non-linear behavior of $\rho_H(B)$ in bismuth occurs up to 200 K, at least. We stress that the hysteresis in $\rho_H(B)$ shown in Fig. 12 cannot be understood using classical approaches [17, 18].

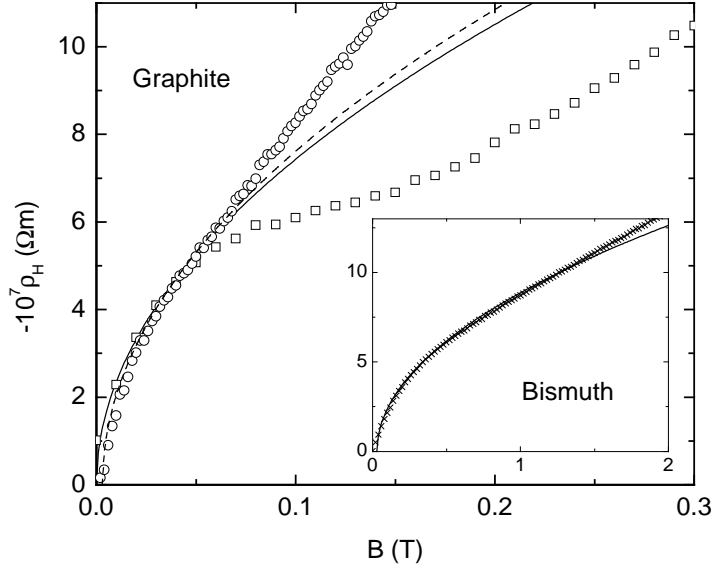

Figure 11. Hall resistivity $-\rho_H(B)$ measured for HOPG-1 ( ) and HOPG-3 (o) samples at T = 5 K. Solid and dashed lines correspond to $-\rho_H(B) \sim B^{1/2}$ and $-\rho_H(B) \sim (B - B_{01})^{1/2}$, $B_{01}$= 0.003T dependencies, respectively. Inset shows $-\rho_H(B)$ measured for Bi-S1 sample at T = 200 K; solid line corresponds to $-\rho_H(B) \sim (B - B_{02})^{1/2}$ dependency with $B_{02} = 0.03$ T.

At low fields $\rho_H(B)$ can be very well described by $\sim (B - B_0)^{1/2}$ dependence, see Fig. 11, also expected for an excitonic gap $\Delta_{EI}$ vs. B behavior [23, 24]. It is interesting to note that a proportionality between nesting-driven SDW gap and the Hall coefficient $R_H = \rho_H(B)/B$ has been obtained near a QCP [42]. On the other hand, our data suggest that $\rho_H(B) \sim \Delta_{EI}(B)$. One can understand this result assuming the occurrence of a field-induced ferromagnetic magnetization $M_{FM} \sim \Delta_{EI}(B)$ [23].

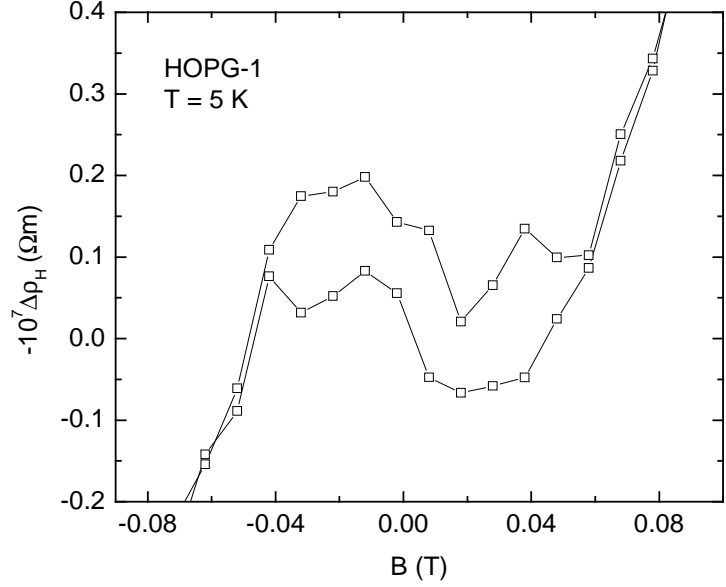

Figure 12. Hall resistivity $-\Delta\rho_H(B)$ obtained for HOPG-1 ( ) at T = 5 K in applied magnetic field of both polarities after subtraction of a fitted polynomial function.

Then, the Hall resistivity can be represented by the sum of two terms [43]

$$\rho_H = R_0 B + R_s \mu_0 M_{FM}, \qquad (2)$$

where $R_0$ and $R_s$ are the ordinary and extraordinary (anomalous) Hall coefficients. If at low fields the second term dominates, $\rho_H(B) \sim M_{FM} \sim \Delta_{EI}(B) \sim (B - B_0)^{1/2}$. The occurrence of ferromagnetism in the doped excitonic insulator state [23] (see also Refs. [44-46]) naturally explains both the non-linearity and the hysteresis in $\rho_H(B)$. Note that due to the huge diamagnetism of both Bi and graphite, an unambiguous detection of $M_{FM}$ by means of magnetization measurements may be a challenging task [47].

In figure 13, we present the results of magnetization measurements performed on bismuth which revealed an enhancement of the diamagnetic signal at T < 7 K measured at low enough fields. This enhancement is effectively suppressed by applied magnetic field of ~ 500 Oe, suggesting that the low-temperature low-field M(T, H) behavior is related to superconductivity [49, 50]. What is particularly interesting is that the suppression of the superconducting signal and MIT take place almost at the same applied magnetic field, see Figs. 2, 7, and 13.

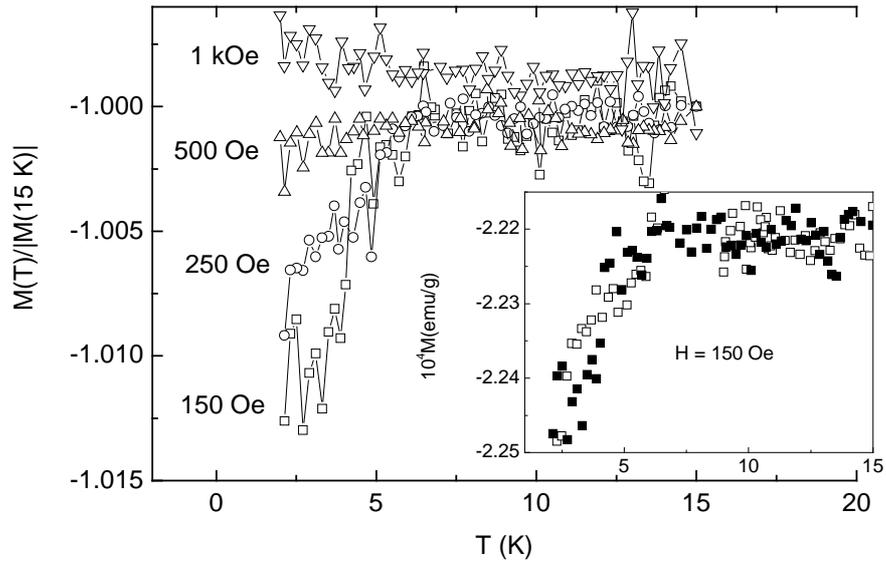

Figure 13. Reduced magnetization $M(T)/|M(T = 15 K)|$ vs. temperature measured for Bi-S3 sample at various applied magnetic fields $H \parallel c$. Inset shows M(T) measured in zero-field-cooled ( ) and field-cooled on cooling (■) regimes with H = 150 Oe.

This fact indicates that the magnetic and transport measuremensts probe essentially the same physical phenomenon. In other words, superconducting correlations are relevant in the analysis of MIT, indeed. Whether T ~ 7 K marks the superconducting transition or just the temperature of visible Meissner effect (with the local superconductivity occurring at higher temperatures within isolated small regions) remains to be elucidated. According to Refs. [49, 50], the superconductivity in granular Bi is associated with "grain" surface of clusters which gives a hint why the 2D BMIT scaling works so well (Fig. 7) in otherwise 3D bismuth samples. It should be emphasized that according to Refs. [25, 26] surface states in Bi are situated very near the Fermi level, leading to reduced effective dimensionality, a fact which has not been previously taken into account.

Thus, based on the above and previously published results [29, 33, 34] we conlude that competing (and possibly coexisting) excitonic and superconducting correlations govern the low-field quantum critical behavior in both graphite and bismuth.

Next, we turn out to the field-induced reentrant IMT. There exists an experimental evidence that the applied magnetic field increases the free carrier density in graphite [51, 52]. This weakens the tendency of the electron-hole liquid to the excitonic instability [23, 24, 53], and thus leads to the reentrant metallic behavior. In agreement with such a scenario, the IMT in graphite is accompanied by the onset of pronounced quantum oscillations [16, 29] also observed in bismuth. In this regime, the Landau level quantization plays a major role, leading in quasi-2D and 2D graphite samples to the quantum Hall effect [16, 19-21]. Is there any relationship between the QHE occurring in quasi-2D graphite and the reentrant metallic phase observed in 3D graphite samples ? A positive answer to this question was already given in Ref. [16], where a cascade of insulator-metal-insulator (I-M-I) transitions associated with the QHE has been found. The same phenomenon has been reported in Ref. [21] for graphene.

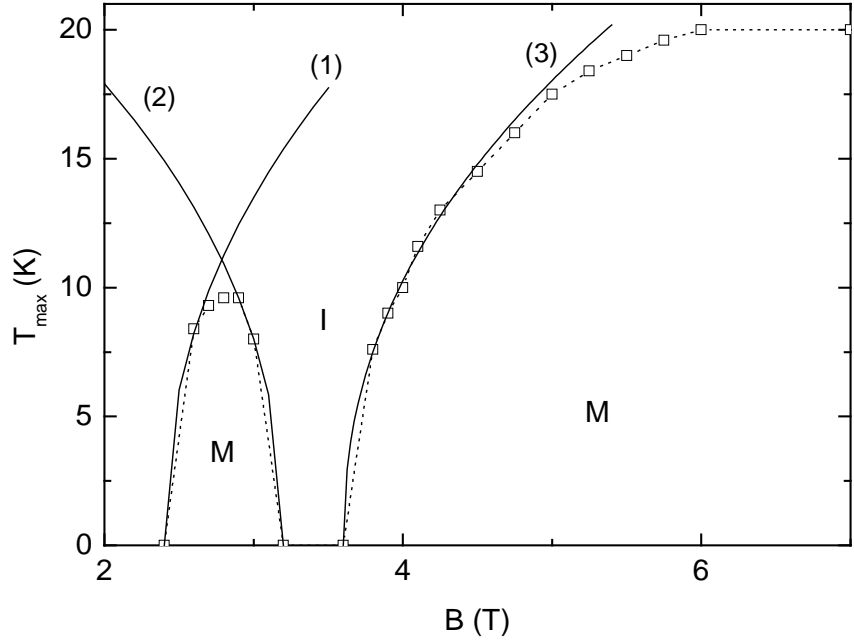

Figure 14. $T_{max}(B)$ measured for HOPG-1 sample [16]; solid lines corresponds to $T \sim (B - B_{c1})^{0.45}$ (curve 1), $T \sim (B_{c2} - B)^{0.45}$ (curve 2), and $T \sim (B - B_{c3})^{0.45}$ (curve 3) dependencies, where $B_{c1} = 2.4$ T, $B_{c2} = 3.2$ T, and $B_{c3} = 3.6$ T (see text).

Here we would like to emphasize this relationship demonstrating $T_{max}$ vs. B behavior for the graphite sample with an intermediate anisotropy $\rho_c/\rho_b \leq 10^4$ (HOPG-1) where no QHE was detected. Figure 14 shows that at high magnetic fields there occur two "metallic" phases; one in the field interval $2.4\ T \leq B \leq 3.2$ T and another for $B \geq 3.6$ T, as well as the intervening "insulating" state at $3.2\ T \leq B \leq 3.6$ T. The I-M and M-I transition boundaries follow the same power-law behavior as given in Fig. 4, i. e. $T_{max}(B) \sim |B - B_c|^{0.45 \pm 0.05}$. We recall that transitions between quantum Hall (QH) plateaus measured in strongly anisotropic samples (HOPG-UC and HOPG-3) [16] revealed a quantum critical behavior with the same exponent κ ~ 0.45.

The universality of the exponent $\kappa \approx 1/\nu z = 0.45 \pm 0.05$ suggests that all metal-insulator and insulator-metal transformations reported in this work belong to the same universality class. Overall the results indicate that this should involve both BM (or superconductor) – excitonic insulator transition and transitions between quantum Hall (QH) plateaus.

In summary, we reported on possible quantum phase transitions which govern the physics of both graphite and bismuth in a broad temperature and applied magnetic field range. A new experimental evidence for the Bose metallic state in which both superconducting and excitonic correlations play a role is obtained. The anomalous Hall effect is measured and atributted to a field-driven ferromagnetic moment induced in the doped excitonic insulator state. We also emphasized an intimate coupling of the reentrant insulator-metal transition in 3D samples and the quantum Hall transitions measured in strongly anisotropic (quasi-2D) graphite.

We thank I. A. Luk'yanchuk, P. Esquinazi, S. G. Sharapov, V. A. Miransky, D. V. Khveshchenko, Shan-Wen Tsai, and D. Das for useful discussions. This work was supported by FAPESP and CNPq.